\documentstyle[prl,epsf,epsfig,amssymb,amsbsy,amstext,amsfonts,aps]{revtex}
\begin{document}
\twocolumn
\bibliographystyle{try}

\topmargin 0.1cm

\wideabs{

\draft{SLAC-PUB-8677\\}
\draft{NORDITA-2000/90 HE\\}
\draft{DAPNIA-SPhN-00-72 }


\def\slac{$^{1}$}
\def\jlab{$^2$}
\def\nordita{$^{3}$}
\def\sphn{$^4$}

\title{Photoproduction of charm near threshold}

\author{S. J. Brodsky,\slac\
               E. Chudakov,\jlab\
               P. Hoyer,\nordita\
               J.M. Laget,\sphn\
}


\address{
\slac Stanford Linear Accelerator Center, Stanford University, Stanford, CA94309, U.S.A.\\
\jlab Thomas Jefferson National Accelerator Facility, Newport News,
VA 23606, USA\\
\nordita NORDITA, Copenhagen, Denmark\\
\sphn CEA-Saclay, DAPNIA-SPhN, F91191 Gif-sur-Yvette Cedex, France (E-mail: jlaget@cea.fr)\\
}
\date{\today}
\maketitle

\begin{abstract}
Charm and bottom production near threshold is sensitive to
the multi-quark, gluonic, and hidden-color
correlations of hadronic and nuclear wavefunctions in QCD
since all of the target's constituents must act coherently
within the small interaction
volume of the heavy quark production subprocess.
Although such multi-parton subprocess cross sections
are suppressed by powers of $1/m^2_Q$, they have less
phase-space suppression and can
dominate the contributions of the leading-twist single-gluon
subprocesses in the threshold regime.  The small rates for
open and hidden charm photoproduction at threshold call for a dedicated
facility.
\end{abstract}

}

\narrowtext

\begin{figure}[h]
\begin{center}
   \leavevmode
   \epsfxsize=5cm
    \epsfbox{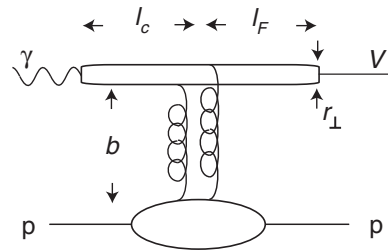}
\end{center}
\caption{ 
The characteristic scales in elastic $J/\psi$ production on protons near threshold,
$E^{\rm lab}_{\gamma}= 8.20$ GeV.
The longitudinal coherence length of the
$c\overline{c}$ fluctuation of the photon
is short, $l_c \protect\cong 2E^{\rm lab}_{\gamma}/4m^2_c = 0.36$ fm. 
The large mass of
the charmed quark also imposes a small
transverse size $r_{\perp}\sim 1/m_c= 0.13$ fm on this fluctuation. 
The minimum momentum transfer is
large, 
$t_{min}\sim - 1.7$ GeV$^2$.  All of the partons of the target
wavefunction have to transfer their energy to the charm quarks within
their proper creation time
$1/m_c$, and must be within this transverse distance from the
$c\overline{c}$ and from each other, so that charm production near 
threshold occurs at small impact distances 
$b \sim 1/m_c \sim 0.13$ fm.
}
\label{time}
\end{figure}

The threshold regime of charmonium and open charm production can provide
a new window into
multi-quark, gluonic, and hidden-color
correlations of hadronic and nuclear wavefunctions in QCD.
For example, consider charm photoproduction
$\gamma ~ p \to J/\psi ~ p$ at the threshold energy $E^{\rm lab}_{\gamma}=
8.20$ GeV.  [See Fig.~\ref {time}.] The available production
energy cannot be wasted at threshold, so all three valence quarks of the
target nucleon must interact coherently within the small interaction
volume of the heavy quark production subprocess.
In the case of threshold charm photoproduction on a deuteron
$\gamma ~ d \to J/\psi ~ d,$ all color configurations of the six valence
quarks will be involved at the short-distance scale $1/m_c$.  Thus the
exchanged gluons can couple to a color-octet quark cluster and reveal the
``hidden-color" part of the nuclear wave function, a domain of short-range
nuclear physics where nucleons lose their
identity \cite{Matveev:1977xt,Matveev:1978ha,Brodsky:1983vf}.

At high energies the dominant contribution
to an inclusive process involving a hard scale $Q$ comes from
``leading twist'' diagrams, characterized by only one parton from each
colliding particle participating in the large momentum subprocess.  Since
the transverse size scale of the hard collision is
$1/Q$, only partons within this distance can affect the
process.  The likelihood that two partons of the incident hadrons can be
found so close to each other is typically proportional to the transverse
area
$1/Q^2$ and leads to the suppression of higher-twist, multi-parton
contributions.
However, in contrast to charm production
at high energy,  charm production near threshold requires all of the
target's constituents to act coherently in the
heavy quark production process: only
compact proton Fock states with a radius of order of the Compton
wavelength of the heavy quark can contribute to charm production at
threshold.
Although the higher-twist subprocess cross sections
are suppressed by powers of $1/m^2_c$, they have much less
phase-space suppression at threshold.  Thus charm production
at threshold is sensitive to short-range correlations between the valence
quarks of the target, and higher-twist multi-gluon exchange reactions can
dominate over the contributions of the leading-twist
single-gluon subprocesses.

One can determine the power-law dependence of multi-parton
heavy quark production subprocesses using an operator product
analysis of the effective heavy quark theory.  The heavy quark
photoproduction cross section can be computed through the optical theorem
from the corresponding cut diagrams of the forward Compton amplitude.
Such diagrams factorize into the convolution of two factors: a heavy
quark loop diagram connecting the photons to the exchanged gluons,
times the gauge invariant matrix element of a product of
gluon field strengths
$<p \vert G^n_{\mu \nu} \vert p>$. Because of the non-Abelian coupling,
a single field strength can correspond to one or two exchanged gluons. For
heavy quark masses,
$m_Q^2
\gg \Lambda^2_{QCD}$ the heavy quark loop
contracts to an effective local operator, so that the field strengths
in the matrix element are all evaluated at the same local point. 
The minimal gluon exchange contribution  ($n=2)$ gives the
leading twist photon-gluon fusion contribution.  
Since $<p \vert G^n \vert p>$ scales as
${(\Lambda^2_{QCD})}^{n-1}$, each extra gluon field strength  
connecting to the heavy quark loop must give a factor of $(1/m^2_Q)$.  
(Higher derivatives
in the matrix element are further suppressed.)
Thus one pays a penalty of a factor $(\Lambda^2/m^2_Q)$ as the number of
exchanged gluon fields is increased.
However, as we shall see,
the suppression from the multiple gluon exchange contributions are
systematically compensated by fewer powers of energy threshold factors,
so that at threshold multi-gluon contributions will dominate.  A similar
effective field theory operator analysis has been used\cite{Franz:2000ee}
to estimate the momentum fraction carried by intrinsic heavy quarks in the
proton \cite{Brodsky:1980pb,Harris:1996jx}.  

In this paper, we will use reasonable conjectures for the short distance
behavior of hadronic matter inferred from properties of perturbative
QCD and effective heavy quark field theory to estimate the behavior of the
reaction cross section.

\begin{figure}
\begin{center}
    \leavevmode
    \epsfxsize=8cm
    \epsfbox{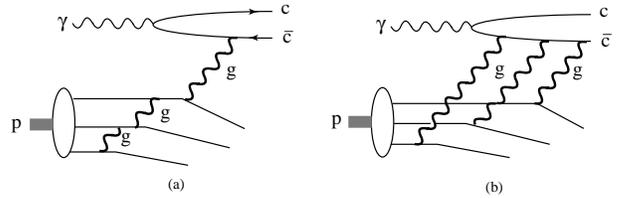}
\end{center}
\caption{\label{graphs} Two mechanisms for transferring most of the proton
momentum to the charm quark pair in $\gamma p \rightarrow c\overline{c}p$ near
threshold.  The leading twist contribution (a) dominates at high energies, but
becomes comparable to the higher-twist contribution (b) close to
threshold.}
\end{figure}

The effective proton radius in charm photoproduction near threshold can be
determined from the following argument~\cite{Ho97,Br92}.  As indicated
in Fig.~\ref{graphs}a,
most of the proton momentum may first be
transferred to one (valence) quark, followed by a hard subprocess $\gamma
q\rightarrow c\overline{c}q$.  If the photon energy is $E_{\gamma}=\zeta
E_{\gamma}^{th}$, where $E_{\gamma}^{th}$ is the energy at kinematic threshold
($\zeta \geq 1$), the valence quark must carry a fraction $x=1/\zeta$ of the
proton (light-cone) momentum.  The lifetime of such a Fock state (in the
light-cone or infinite momentum frame) is $\tau = 1/\Delta E$, where
\begin{equation}
\Delta E= \frac{1}{2p}\left[ m_p^2 -\sum_{i}^{} \frac{p_{i\perp}^2 +
m_i^2}{x_i}\right] \simeq \frac{\Lambda_{QCD}^2}{2p(1-x)}
\end{equation}
For $x=1/\zeta$ close to unity such a short lived fluctuation can be created
(as indicated in Fig.~\ref{graphs}a) through momentum transfers from valence
states (where the momentum is divided evenly) having commensurate lifetimes
$\tau$ and transverse extension
\begin{equation}
r_{\perp}^2\simeq \frac{1}{p_{\perp}^2} \simeq\frac{\zeta -1}{\Lambda_{QCD}^2}
\end{equation}
This effective proton size thus decreases towards threshold ($\zeta \rightarrow
1$), reaching $r_{\perp}^2 \simeq 1/m_c^2$ at threshold ($\zeta -1 \simeq
\Lambda_{QCD}^2/m_c^2$).

As the lifetimes of the contributing Fock states approach the time scale of the
$c\overline{c}$ creation process, the time ordering of the gluon exchanges
implied by Fig.~\ref{graphs}a ceases to dominate higher-twist
contributions such as that of Fig.~\ref{graphs}b~\cite{Br92}.  There are
in fact reasons to expect that the latter diagrams give a dominant
contribution to charmonium production near threshold.  First, there are
many more such diagrams.  Second, they allow the final state proton to
have a small transverse momentum (the gluons need $p_{\perp}\simeq m_c$
to couple effectively to the $c\overline{c}$ pair, yet the overall
transfer can still be small in Fig.~\ref{graphs}b). Third, with several
gluons coupling to the charm quark pair its quantum numbers can match
those of a given charmonium state without extra gluon emission.

The above discussion is generic, and does not indicate how close to threshold
the new effects actually manifest themselves.  While this question can only be
settled by experiment, we rely on a simple model to get an estimate of the
cross section.

Near-threshold charm production probes the $x\simeq 1$ configuration in
the target, the spectator partons carrying a vanishing fraction $x\simeq
0$ of the target momentum.  This implies that the production rate behaves
near
$x\rightarrow 1$ as $(1-x)^{2n_s}$ where $n_s$ is
the number of
spectators \cite{Brodsky:1995kg}.  
Perturbative QCD predicts three different gluonic components of the
photoproduction cross-section: i) The leading twist $(1-x)^4$ distribution for
the process $\gamma q\rightarrow c \overline{c} q$, which leaves two quarks
spectators (Fig.~\ref{graphs}a); ii) Scattering on two quarks in the proton
with a net distribution $\frac{(1-x)^2}{R^2{\cal{M}}^2}$, $\gamma qq
\rightarrow c\overline{c} qq$, leaving one quark spectator; iii) Scattering on
three quark cluster (Fig.~\ref{graphs}b) in the proton with a net distribution
$\frac{(1-x)^0}{R^4{\cal{M}}^4}$, $\gamma qqq \rightarrow c\overline{c} qqq$,
leaving no quark spectators.
There is some arbitrariness in the definition of $x$ close to threshold. We shall use  $x = ({2m_p\cal{M}+\cal{M}}^2)/(s-m_p^2)$, 
where $s=E_{CM}^2$ and $\cal{M}$ is the mass of the $c\overline{c}$ pair,
which has the property $x=1$ at threshold.
The relative weight of
scattering from multiple quarks is given by the probability $1/R^2{\cal{M}}^2$
that a quark in the proton of radius $R\simeq 1$~fm is found within a
transverse distance $1/\cal{M}$ (see Ref.~\cite{Br79}).

The two-gluon exchange contribution produces odd $C$ quarkonium $\gamma
gg\rightarrow J/\psi$, thus permitting exclusive $\gamma p \rightarrow
J/\psi p$ production.  The photon three-gluon coupling $\gamma ggg
\rightarrow c\overline{c}$ produces a roughly constant term at threshold
in
$\sigma/v$, where it is expected to dominate (here $v=1/16\pi
(s-m_p^2)^2$   
is the usual phase space factor).  It produces the $\eta_cp$, $\chi_cp$
and other C even resonances, but also
$J/\psi$.

For elastic charm production (when the proton target remains bound), it is also
necessary to take into account the recombination of the three valence quarks
into the proton via its form factor, as well as the coupling of the photon to
the $c\overline{c}$ pair.  For two gluon exchange the cross section of the
$\gamma p \rightarrow J/\psi p$ takes the form:
\begin{equation}
\frac{d\sigma}{dt}={\cal N}_{2g}v \frac{(1-x)^2}{R^2{\cal{M}}^2}
F^2_{2g}(t) (s-m_p^2)^2   
\label{eq3}
\end{equation}
while for three gluon exchange it takes the form:
\begin{equation}
\frac{d\sigma}{dt}={\cal N}_{3g}v \frac{(1-x)^0}{R^4{\cal{M}}^4}
F^2_{3g}(t) (s-m_p^2)^2   
\label{eq4}
\end{equation}
where $F_{2g}(t)$ and $F_{3g}(t)$ are proton form factors that
take into account the fact that the three target quarks recombine 
into the final proton after the emission of two or three gluons. 
While they
are analogous to  
the proton elastic form factor $F_1(t)$, they are not
known.  
In the numerical applications, we have parameterized them as  
$F^2=\exp(1.13 t)$, according to the experimental $t$ dependency
of the cross section~\cite{Gi75}.  The $(s-m_p^2)^2$  
term comes from the coupling of the incoming photon to the
$c\overline{c}$ pair and the spin-$1$ nature of gluon exchange (see,
for instance, Ref.~\cite{La00}).  It compensates the same term in the
phase space $v$.  The normalization coefficient ${\cal N}$ is
determined assuming that each channel saturates the experimental
cross section measured at SLAC~\cite{Ca75} and Cornell~\cite{Gi75}
around $E_{\gamma} = 12$ GeV.

\begin{figure}
\begin{center}
\epsfig{width=6.5cm,file=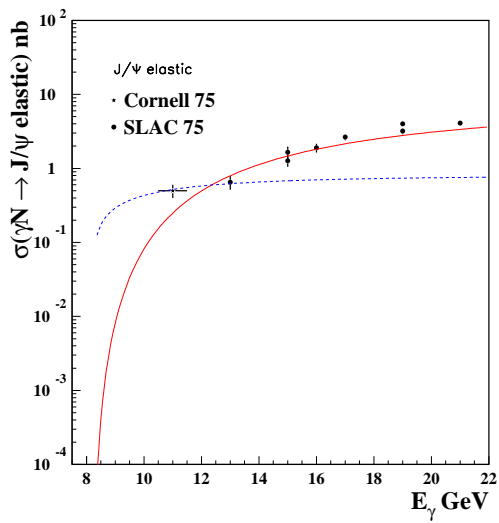}
\end{center}
\caption{\label{X-sections} Variation of the
J/$\psi$ photoproduction cross section  
near threshold. Solid line:  
two gluon exchange (Eqs.~\ref{eq3}). Dashed line: three gluon 
exchange (Eq.~\ref{eq4}).}
\end{figure}

Notice that expressions~(\ref{eq3}) and~(\ref{eq4}) are valid in a limited
energy range near threshold, where $x\sim 1$.  To be more specific, $x=0.82$ at $E^{\rm lab}_{\gamma}= 10$~GeV and $x=0.69$ at $E^{\rm lab}_{\gamma}= 12$~GeV. So we expect that our model still makes sense up to the lowest energy range where experimental data exist.
At higher energies  
one has to
rely on the variation of the gluon distribution in the vicinity
of $x\sim 0$ to
reproduce the steep rise of charm photoproduction~\cite{BrXX,Kh98} above $E^{\rm lab}_{\gamma}\approx  100$~GeV ($x\leq 0.082$). 

As shown in Fig.~\ref{X-sections},
the threshold dependence of our conjectured
cross sections~(\ref{eq3}) and~(\ref{eq4}) is consistent with the scarce
existing
data~\cite{Gi75,Ca75}.  
Indeed, there is also evidence~\cite{An77}
that the energy dependence of the $J/\psi$ elastic photoproduction cross
section at forward angles is roughly flat up to $E_{\gamma} \approx 12$ GeV,
in contrast to the steep variation observed at higher energies.  More
accurate measurements of the J/$\psi$ elastic photoproduction cross
section up to about 20 GeV are clearly needed.

The existence of five-quark resonances near threshold in the $\gamma
p\rightarrow p c\bar c$ process~\cite{Br90} would modify our picture.  However,
the qualitative features of the two- and three-gluon-exchange cross
sections (which differ by orders of magnitude near threshold) should
remain valid.

\begin{figure}
\begin{center}
\leavevmode
\epsfxsize=5cm
\epsfbox{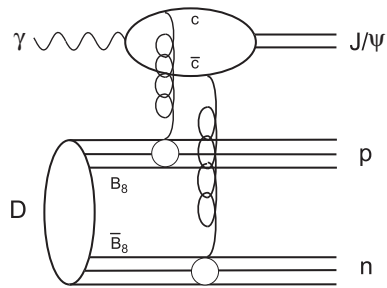}
\end{center}
\caption{\label{hidden_col} The simplest diagram which reveals a
hidden-color state in deuterium~\protect\cite{La94}.}
\end{figure}

On few body targets, each exchanged gluon may couple to a colored quark cluster
and reveal the hidden-color part of the nuclear wave function, a domain of
short-range nuclear physics where nucleons lose their identity.  The
existence of such hidden-color configurations is predicted by QCD
evolution equations~\cite{Brodsky:1983vf}.  It is striking that in $\gamma
d\rightarrow J/\psi pn$, (Fig.~\ref{hidden_col}), the
$|B_8\overline{B_8}>$ hidden-color state of the deuteron couples so
naturally via  
two gluons to the $J/\psi pn$ final state~\cite{La94},
since the coupling of a single gluon to a three-quark cluster turns it
from a color octet to a singlet.

\begin{figure}
\begin{center}
    \leavevmode
    \epsfxsize=6cm
    \epsffile{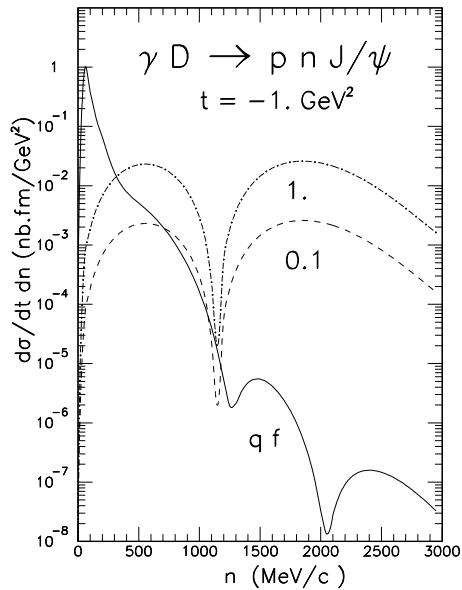}
\end{center}
\caption{\label{d_psi} The variation of the cross-section of the reaction
$\gamma D\rightarrow pn J/\psi$ against the neutron momentum $|\vec{n}|$, at
fixed $t$.  Solid line: quasi-free contribution.
Dashed line: contribution of a
hidden-color component when its probability is 0.1\%.  Dash-dotted curve:
the same for a probability of
1\%.}  
\end{figure}

When the nucleon is embedded in a
nuclear medium, two mechanisms
govern the photo- and
electroproduction of $J/\psi $ mesons.  The first,
the quasi-free production mechanism, contributes the following cross section to
the $\gamma d \rightarrow J/\psi pn$ reaction, when integrated over the angles
of the spectator neutron~\cite{La81}:
\begin{eqnarray}
\frac{d\sigma}{dt d\!\mid\vec{n}\mid}=
\left. \frac{d\sigma}{dt}\right|_{\gamma p\rightarrow J/\psi p}
4\pi \vec{n}^2\rho (\mid \vec{n} \mid) \\
\int\rho (\mid \vec{n} \mid) d\vec{n}=1
\end{eqnarray}
where $\mid\vec{n}\mid$ is the momentum of the spectator neutron.  The nucleon
momentum distribution
$\rho (\mid \vec{n} \mid)$ in deuterium decreases very quickly~\cite{La81}
with increasing neutron momentum.  Consequently, by selecting high neutron
momenta one can suppress quasi-free production and measure inherently nuclear
effects.  The quasi-free contribution in Fig.~\ref{d_psi} has been computed
with the Paris wave function~\cite{Pa81} of the deuterium.

The second contribution comes from coupling
the two gluons to
separate color octet 3-quark clusters.  This contribution
is expected to have a
flatter momentum distribution, since the recoil momentum is shared
between
two nucleons.  The corresponding cross section can be roughly estimated as~:
\begin{equation}
\frac{d\sigma}{dtd\!\mid\vec{n}\mid}=     \left.
\frac{d\sigma}{dt}\right|_{\gamma    p\rightarrow    J/\psi    p}
4\pi \vec{n}^2\left[
\varphi_{cc}       (\frac{\vec{n}}{2})       \right]^2       \frac{F_1^4
(\frac{t}{4})}{F_1^2 (t)} \label{ampd}
\end{equation}
where the fourth power of the nucleon form factor comes from the fact that
two  nucleons have to recombine,  each at the momentum transfer
$t/4$ \cite{La94,Brodsky:1983kb}.  
We assume that the form factor of the transition between a colored
cluster and the nucleon does not differ too much from the nucleon form
factor~\cite{note} and
that the recoil momentum is equally
shared between the two colored clusters whose relative wave
function is $\varphi_{cc} (\frac{\vec{n}}{2})$.  This component of the
deuterium wave function has not been measured
and
few predictions are available.  As
an example and to set the order of magnitude, the hidden-color
contribution in Fig.~\ref{d_psi} has been obtained
using the Fourier transform of the
wave function depicted in Fig.~11  of  Ref.~\cite{Ya86}.  Since it exhibits
a node around 500 MeV/c,  a node appears in the cross-section around $n\simeq
1\;GeV/c$.
In a more elaborate calculation the sum over
the nucleon internal momentum would wash out this node.
Anyway,  this rough estimate shows that the hidden-color component
contribution dominates the cross-section above 0.5 GeV/c.  The calculation
reported in~\cite{Ya86} predicts a probability of finding a hidden-color
component in the deuterium wave function of the order of $0.1\%$.
Fig.~\ref{d_psi} also shows what one may expect for a probability
around $1\%$.

Scattering on colored clusters may dominate subthreshold production, since the
high momentum of the struck
nucleon suppresses the quasi free mechanism.
On deuterium the
threshold for $J/\psi$ production is $\sim 5.65$ GeV, while on heavy nuclei the
threshold is simply the $J/\psi$ mass $3.1$ GeV.

Let us close this note with two remarks.  At threshold, the formation length~\cite{Fa90,Ko91}
(during which the $c\overline{c}$ pair evolves into a $J/\psi$, after its
interaction with a nucleon)
\begin{equation}
l_F\cong \frac{2}{m_{\psi'}-m_{J/\psi}}\left[\frac{E_{J/\psi}}{2m_c} \right]
\cong 0.22 {\rm\ fm\ } E_{\gamma} {\rm /GeV}  
\end{equation}
  is around 1 fm, considerably smaller than the size of a large nucleus.
It is thus possible to determine the scattering cross
section of a full sized
charmonium  
on a nucleon
using nuclear targets, in contrast to the situation at
higher energies where
the nuclear interaction of a compact
$c\overline{c}$ pair is measured.
The study of the A dependence of the J/$\psi$
photoproduction cross section at SLAC at 20 GeV~\cite{An77b} gave
$\sigma_{J/\psi N}=3.5\pm 0.8\pm 0.5$ mb.  However, a large calculated background
was subtracted and the lack of information on the J/$\psi$ kinematics
prevented a separate measurement of
coherent and incoherent photoproduction.
A new measurement of J/$\psi$ photoproduction on several nuclei
around
$E_{\gamma}\approx 10$ GeV, with good particle
identification and a determination of
the J/$\psi$ momentum is clearly called for.

Although the $c\overline{c}$ pair is created with rather high momentum
at threshold, it may be possible to observe reactions where the pair
is captured
by the target nucleus, forming ``nuclear-bound quarkonium''~\cite{Br90}.
This process should be
enhanced  
in subthreshold reactions.  There is no Pauli blocking
for charm quarks in nuclei, and it has been estimated that there is a large
attractive Van der Waals potential binding the pair to the nucleus~\cite{Lu92}.
The discovery of such qualitatively new states of matter would be
very important.

In this paper we have shown that charm production
near threshold has strong sensitivity to
the multi-quark, gluonic, and hidden-color
correlations of hadronic and nuclear wavefunctions in QCD.
Although multi-parton subprocess cross sections
are suppressed by powers of $1/m^2_c$, they have
correspondingly  
less phase-space suppression and thus can
dominate the contributions of the leading-twist single-gluon
subprocesses in the threshold regime.
Such processes will
therefore add to our understanding of the short-range structure of
nucleons and nuclei.  A new dedicated facility, such as CEBAF12 or
ELFE, with high intensity and duty factor,  will make possible the
experiments needed to explore this QCD frontier.

We acknowledge a simulating discussion with M. Strikman.
This work has been supported in part by the Department of Energy
under contracts  DE-AC03-76SF00515(SB),
DE-AC05-84ER40150 (EC) and the European Commission under contract
FRMX-CT-96-0008 (PH and JML).


\begin{thebibliography}{99}
\bibitem{note}  This is a conservative estimate, since a color
3-quark cluster is probably  more compact than a singlet nucleon;
hence the form factor should be larger.
\end{thebibliography}
\end{document}